\documentclass[acmsmall]{acmart}

\usepackage{xspace}
\usepackage{xcolor}
\usepackage{soul}

\usepackage{graphicx}
\usepackage{multirow}
\usepackage{bbding}
\usepackage{hyperref}

\usepackage{colortbl}

\usepackage{bigstrut}
\usepackage{soul}
\usepackage[most]{tcolorbox}
\usepackage{multirow}
\usepackage{tabularx}

\usepackage{bbding}
\usepackage{hyperref}

\AtBeginDocument{%
  }

\setcopyright{acmlicensed}
\copyrightyear{2025}
\acmYear{2025}
\acmDOI{XXXXXXX.XXXXXXX}

\acmConference[GROUP '25]{The 2025 ACM International Conference on Supporting Group Work}{January 12--15,  2025}{Hilton Head, SC}
\acmISBN{978-1-4503-XXXX-X/18/06}

\begin{document}

\title[Exploring Large Language Models Through a Neurodivergent Lens]{Exploring Large Language Models Through a Neurodivergent Lens: Use, Challenges, Community-Driven Workarounds, and Concerns}

\author{Buse Carik}
\orcid{0000-0002-4511-5827}
\email{buse@vt.edu}
\affiliation{%
  \institution{Virginia Tech}
  \city{Blacksburg}
  \state{Virginia}
  \country{USA}
}

\author{Kaike Ping}
\email{pkk@vt.edu}
\affiliation{%
  \institution{Virginia Tech}
  \city{Blacksburg}
  \state{Virginia}
  \country{USA}
}

\author{Xiaohan Ding}
\email{xiaohan@vt.edu}
\affiliation{%
  \institution{Virginia Tech}
  \city{Blacksburg}
  \state{Virginia}
  \country{USA}
}

\author{Eugenia H. Rho}
\email{eugenia@vt.edu}
\affiliation{%
  \institution{Virginia Tech}
  \city{Blacksburg}
  \state{Virginia}
  \country{USA}
}

\renewcommand{\shortauthors}{Carik et al.}
\definecolor{buse_skyblue}{rgb}{0.1098, 0.5686, 0.7411}
\definecolor{buse_orange}{rgb}{0.9450, 0.6117, 0}
\definecolor{buse_plum}{rgb}{0.7, 0.3, 0.5}
\definecolor{buse_green}{rgb}{0.0274, 0.4274, 0.2352}
\definecolor{buse_mustard}{rgb}{0.8, 0.7, 0.2} 
\definecolor{buse_red}{rgb}{0.8392, 0.3921, 0.2666}
\definecolor{buse_purple}{rgb}{0.6, 0.4, 0.7} 
\definecolor{buse_seafoam}{rgb}{0.5, 0.75, 0.6}
\definecolor{buse_blue}{rgb}{0.3333, 0.3764, 0.6627}

\newcommand{\textft}[1]{#1}
\newcommand{\customul}[2][black]{\setulcolor{#1}\ul{#2}\setulcolor{black}}

\newcommand{\EMt}[1]{\emph{\textcolor{buse_skyblue}{#1}}}
\newcommand{\MHt}[1]{\emph{\textcolor{buse_purple}{#1}}}
\newcommand{\IPt}[1]{\emph{\textcolor{buse_green}{#1}}}
\newcommand{\LRt}[1]{\emph{\textcolor{buse_orange}{#1}}}
\newcommand{\PRt}[1]{\emph{\textcolor{buse_plum}{#1}}}
\newcommand{\Challenget}[1]{\emph{\textcolor{buse_red}{#1}}}
\newcommand{\Needst}[1]{\emph{\textcolor{buse_seafoam}{#1}}}
\newcommand{\Sharet}[1]{\emph{\textcolor{buse_mustard}{#1}}}
\newcommand{\Concernt}[1]{\emph{\textcolor{buse_blue}{#1}}}

\newcommand{\EMcode}[1]{{\textft{\customul[buse_skyblue]{#1}}}}
\newcommand{\MHcode}[1]{{\textft{\customul[buse_purple]{#1}}}}
\newcommand{\IPcode}[1]{{\textft{\customul[buse_green]{#1}}}}
\newcommand{\LRcode}[1]{{\textft{\customul[buse_orange]{#1}}}}
\newcommand{\PRcode}[1]{{\textft{\customul[buse_plum]{#1}}}}
\newcommand{\Challengecode}[1]{{\textft{\customul[buse_red]{#1}}}}
\newcommand{\Needscode}[1]{{\textft{\customul[buse_seafoam]{#1}}}}
\newcommand{\Sharecode}[1]{{\textft{\customul[buse_mustard]{#1}}}}
\newcommand{\Concerncode}[1]{{\textft{\customul[buse_blue]{#1}}}}

\newtcbox{\ES}{on line,
  arc=3pt,outer arc=3pt,
  colback=buse_skyblue!25!white, colframe=blue!50!black,
  boxsep=0pt,left=2pt,right=2pt,top=2pt,bottom=2pt,
  boxrule=0pt}

\newtcbox{\MH}{on line,
  arc=3pt, outer arc=3pt,
  colback=buse_purple!25!white, colframe=pink!50!black,
  boxsep=0pt, left=2pt, right=2pt, top=2pt, bottom=2pt,
  boxrule=0pt  
}

\newtcbox{\IP}{on line,
  arc=3pt,outer arc=3pt,
  colback=buse_green!25!white, colframe=green!50!black,
  boxsep=0pt,left=2pt,right=2pt,top=2pt,bottom=2pt,
  boxrule=0pt}

\newtcbox{\IPP}{on line,
  arc=3pt,outer arc=3pt,
  colback=buse_green!25!white, colframe=green!50!black,
  boxsep=0pt,left=2pt,right=2pt,top=2pt,bottom=4pt,
  boxrule=0pt}

\newtcbox{\LR}{on line,
  arc=3pt,outer arc=3pt,
  colback=buse_orange!25!white, colframe=orange!50!black,
  boxsep=0pt,left=2pt,right=2pt,top=2pt,bottom=2pt,
  boxrule=0pt}

\newtcbox{\PR}{on line,
  arc=3pt,outer arc=3pt,
  colback=buse_plum!25!white, colframe=purple!50!black,
  boxsep=0pt,left=2pt,right=2pt,top=2pt,bottom=2pt,
  boxrule=0pt}

\newtcbox{\Challenge}{on line,
  arc=3pt,outer arc=3pt,
  colback=buse_red!25!white, colframe=pink!50!black,
  boxsep=0pt,left=2pt,right=2pt,top=2pt,bottom=2pt,
  boxrule=0pt}

\newtcbox{\Needs}{on line,
  arc=3pt,outer arc=3pt,
  colback=buse_seafoam!25!white, colframe=green!50!black,
  boxsep=0pt,left=2pt,right=2pt,top=2pt,bottom=4pt,
  boxrule=0pt}

\newtcbox{\Share}{on line,
  arc=3pt,outer arc=3pt,
  colback=buse_mustard!25!white, colframe=yellow!50!black,
  boxsep=0pt,left=2pt,right=2pt,top=2pt,bottom=2pt,
  boxrule=0pt}

\newtcbox{\Concern}{on line,
  arc=3pt,outer arc=3pt,
  colback=buse_blue!25!white, colframe=blue!50!black,
  boxsep=0pt,left=2pt,right=2pt,top=2pt,bottom=4pt,
  boxrule=0pt}
\begin{abstract}
Despite the increasing use of large language models (LLMs) in everyday life among neurodivergent individuals, our knowledge of how they engage with, and perceive LLMs remains limited. In this study, we investigate how neurodivergent individuals interact with LLMs by qualitatively analyzing topically related discussions from 61 neurodivergent communities on Reddit. Our findings reveal 20 specific LLM use cases across five core thematic areas of use among neurodivergent users: emotional well-being, mental health support, interpersonal communication, learning, and professional development and productivity. We also identified key challenges, including overly neurotypical LLM responses and the limitations of text-based interactions. In response to such challenges, some users actively seek advice by sharing input prompts and corresponding LLM responses. Others develop workarounds by experimenting and modifying prompts to be more neurodivergent-friendly. Despite these efforts, users have significant concerns around LLM use, including potential overreliance and fear of replacing human connections. Our analysis highlights the need to make LLMs more inclusive for neurodivergent users and implications around how LLM technologies can reinforce unintended consequences and behaviors.
\end{abstract}

\ccsdesc[500]{Human-centered computing~Empirical studies in HCI}
\ccsdesc[500]{Human-centered computing~Empirical studies in accessibility}

\keywords{Large Language Models, Neurodiversity, Autism, Social Anxiety, ADHD, Dyslexia, Reddit}

\maketitle

\section{Introduction}
Neurodiversity refers to variation in the human brain function, behavioral traits, and therefore the human experience of the world \cite{glenn_neuroqueer_2022}. The term \textit{neurodiversity} was originally introduced through the study of online communities of autistic people in the late 90s \cite{singerOddPeopleBirth1998}. Since then, the term has expanded to include dyslexia, dyscalulia, dyspraxia/developmental coordination disorder, Asperger's syndrome, Tourette's syndrome, attention-deficit/hyperactivity disorder (ADHD), and social anxiety \cite{clouder2020neurodiversity}. It is estimated that approximately 15-20\% of the world's population is neurodivergent \cite{DCEG_Staff_2022}. The research community nowadays defines neurodiversity as the diverse range in human cognitive functioning, encompassing unique abilities and ways in which individuals interact with others and their environment \cite{glenn_neuroqueer_2022}. Such variation in cognitive functioning is associated with differences in communicating, reading, writing, and interpreting emotions \cite{LeFevre_2023}. While these differences are increasingly recognized as strengths by society \cite{doyle2020neurodiversity, stenningNeurodiversityStudiesMapping2021}, neurodivergent individuals still face challenges in everyday life, including in areas like employment  \cite{baldwin2014employment,caminiti_jpmorgan_2022}, learning experiences in classrooms \cite{mirfin-veitchRespondingNeurodiversityEducation2020}, and navigating interactions in both professional \cite{krzeminskaAdvantagesChallengesNeurodiversity2019} and personal settings \cite{robertsonNeurodiversityQualityLife2010}. 

As ChatGPT has gained popularity in recent years, many individuals who self-report as neurodivergent are increasingly using large language models (LLMs) to accommodate their unique cognitive processing styles and to receive support with day-to-day tasks \cite{jang2024s,Hoover,stokel-walker_ai_2023}. For instance, consider Billy, a project manager in New Zealand who lives with autism, ADHD, and dyslexia \cite{stokel-walker_ai_2023}. Billy uses LLM-supported tools like Grammarly and ChatGPT to communicate more easily with people around him. To help navigate disagreements with her roommates, one female college student with autism turns to ChatGPT to practice possible conversation scenarios \cite{Hoover}. According to this student, the availability of a conversational partner, though not human, nor perfect, allows her to have a greater sense of independence by not forcing her to rely on her parents all the time in such situations. 

Likewise, there is an increasing number of LLM-supported tools specifically related to neurodiversity or designed for neurodivergent users. For instance, online platforms like Goblin Tools \cite{goblin_2024} aims to help neurodivergent people with tasks they find overwhelming or challenging through a collection of simple, single-tasks guided by LLMs. Others have developed custom ChatGPTs \cite{hustle_playground_neurodiversity,bradley_neurodiversity_navigator} using OpenAI's user-interface features to help both neurotypical and neurodivergent people learn more about topics related to neurodiversity. 
 
While these examples illustrate a facet of how neurodivergent people use LLMs, our understanding of the broader range of LLM use cases, perceptions, challenges, and workarounds within neurodivergent communities remains limited. Our work aims to address this gap by answering the following research questions:

\begin{itemize}
    \item [RQ1.] How are neurodivergent individuals using LLMs?
    \item [RQ2.] When interacting with LLMs, what:
    \begin{itemize}
        \item [(A)] \textbf{\textit{challenges}} do neurodivergent individuals encounter?
        \item [(B)] \textbf{\textit{needs}} and \textbf{\textit{preferences}} do neurodivergent individuals have? 
        \item [(C)] \textbf{\textit{hacks}} and \textbf{\textit{resources}} do neurodivergent individuals share to navigate these challenges?
        \item [(D)] \textit{\textbf{concerns}} do neurodivergent individuals have? 
    \end{itemize}
\end{itemize}

To address our research questions, we conducted a qualitative analysis of online posts from 61 online neurodivergent communities on Reddit related to autism, ADHD, social anxiety, and dyslexia. Our analysis focused on discussions by neurodivergent users about their usage and experiences with LLMs. While we acknowledge that neurodiversity comprises a range of conditions, our work focuses on communities across these four conditions due to their active online presence and significant discussion volumes related to the use of artificial intelligence (AI) and LLM tools. By analyzing this data, our work examines how neurodivergent individuals use LLMs, the challenges they encounter, the workarounds they employ to mitigate limitations, and their preferences and concerns when using LLMs.

Our paper identifies five primary thematic areas in which neurodivergent people use LLMs: \ES{Emotional Well-Being}, \MH{Mental Health Support}, \IP{Interpersonal Communication}, \LR{Learning}, and \PR{Professional Development and Productivity}, including twenty specific use cases across these categories. 

Neurodivergent individuals use LLMs for emotional well-being by engaging with them as a non-judgmental listener or a casual talking buddy, which helps in emotional regulation. For mental health support, many use LLMs to complement traditional therapy to overcome barriers such as financial cost and limited therapy sessions with a professional therapist. Furthermore, many use LLMs to support everyday interpersonal communication. For example, ChatGPT often acts as a neurodivergent-to-neurotypical translator, helping users bridge communication gaps by interpreting social situations or conveying an appropriate tone. For learning, many use LLMs as personal tutors to ask unlimited questions without feeling judged or to enhance their accessibility support. In professional contexts or workplace settings, many use LLMs for task organization, brainstorming, and other career development activities, such as writing cover letters and resumes. 

Our work also highlights key \Challenge{Challenges}, \Needs{Needs and Wants}, \Share{Sharing Hacks \& Resources}, and \Concern{Concerns} related to LLM use across these communities. For example, some users find ChatGPT's responses to be overly neurotypical, failing to capture their unique thought processes. Furthermore, the text-centric nature of LLM-user interactions poses barriers for those with reading and writing difficulties, prompting community discussions around the need and preferences for multimodal capabilities in LLMs. In response to these challenges, some users actively seek advice by sharing prompts and corresponding LLM responses. Others develop workarounds by experimenting and modifying prompts to be more ND-friendly through trial and error. Such users often share their adaptations and hacks with the community to help others benefit as well. Despite these efforts, users have significant concerns around LLM use, including possible over-dependence on AI, which some worry might lead to diminished social skills or making connections with other people.

In summary, we make the following contributions through this work: 

\begin{enumerate}
    \item A comprehensive overview of the diverse thematic areas of LLM use among neurodivergent users and rich characterizations of specific use cases across these areas,
    \item Identification of key challenges, needs, wants, workarounds, and concerns that neurodivergent users encounter when using LLMs,
    \item Exploration of how neurodivergent subreddits serve as discursive communities where individuals can discuss their experiences around using LLMs in everyday life.
\end{enumerate}

\section{Related Work}
\subsection{HCI Technologies for Neurodivergent Users}
Researchers in Human-Computer Interaction (HCI) have developed various technologies to improve the experiences of neurodivergent users in social interactions \cite{escobedo2012mosoco,cha2021exploring,choi_2023}, mental health \cite{bharatharaj2018social,elbeleidy_2022}, learning experiences \cite{yaneva_2015,supangan_2019,gupta_2021}, and interactions at the workplace  \cite{burke2018using}. For example, Choi et al. (2023) investigated the online dating experiences of autistic adults \cite{choi_2023}, while Burke et al. (2018) developed a tool to train autistic individuals to enhance communication skills in job interviews with a virtual character \cite{burke2018using}. Furthermore, another study developed a social media writing support tool for people with dyslexia by converting dyslexic-style writing into standard writing while preserving slang and unique expressions \cite{Wu_Reynolds_Li_Guzmán_2019}. Researchers also developed mobile applications, for instance, to assist children with dyslexia in improving their reading skills \cite{gupta_2021} or to serve as a teaching aid for children with ADHD \cite{supangan_2019}. 

As such, there is an extensive volume of research in HCI focused on developing technologies tailored to support neurodivergent users. Equally important, however, is understanding how individuals use current and emergent technologies in everyday life \cite{daltonNeurodiversityHCI2013}. As highlighted by Spiel et al. (2022), exploring how neurodivergent people customize and interact with existing technologies can reveal significant insights that can contribute to the development of more inclusive technologies in the future \cite{spiel2022adhd}. Therefore, the increasing use of LLMs across neurodivergent communities warrants a deeper investigation into how individuals from these communities are using this emergent technology in their everyday lives. By doing so, we aim to bring forth a more in-depth understanding of the community's diverse perceptions and practices around using LLMs, the challenges they face and associated workarounds, as well as their preferences and concerns in integrating the use of LLMs in their daily lives.

\subsection{LLM Applications for Assisting Neurodivergent Individuals in Daily Tasks}
Recent advancements in language models have led to the growth of scholarship examining the use or application of  LLMs for neurodivergent users. While the majority of these studies have initially concentrated on diagnosing neurodivergent conditions like ADHD and autism \cite{gao_dl_adhd_diag,song2019use,shahamiri2020autism,mukherjee2023detection}, the focus has also expanded into building applications for everyday life. For example, Shen et al. (2022) developed KWickChat, an LLM application enhancing augmentative and alternative communication (AAC) tools for autistic individuals \cite{shen2022kwickchat}. Similarly, Fang (2023) introduced SocializeChat, which leverages GPT-4 to generate tailored conversation suggestions, enhancing social interactions through eye-gaze technology for the autism community \cite{fang_2023}. Mishra and Welch (2024) created a social robot using LLMs to aid children with autism in social understanding through context-specific interactions \cite{mishra2024towards}, while Pai et al. (2024) integrated multimodal interaction to improve therapy sessions for autistic individuals \cite{pai_2024}. Additionally, Goodman et al. (2022) introduced LaMPost, an LLM-powered tool that organizes ideas and improves text clarity and tone to assist dyslexic adults in email communication \cite{Goodman_Buehler_LamPost}.

Despite these advancements, a significant gap remains in understanding how neurodivergent individuals use LLM tools in their daily lives. Prior studies have documented positive experiences for neurodivergent users with LLMs in specific areas, such as workplace communication \cite{jang2024s}, therapy settings \cite{berrezueta2024future,cho2023evaluating}, and reading assistance \cite{tamdjidi2023chatgpt}. For instance, Jang et al. (2024) observed that autistic adults prefer LLM responses to human ones for communication advice in workplace settings due to their non-judgmental nature and immediate availability \cite{jang2024s}. Nevertheless, the variability in the preferences for using LLMs across different social interactions and among various neurodivergent conditions has yet to be fully explored. Furthermore, while studies like that of Berrezueta et al. (2024) demonstrate the potential of using ChatGPT in therapy sessions for individuals with ADHD \cite{berrezueta2024future}, the study focuses on specific therapeutic contexts for ADHD users and overlooks the broader application for other neurodivergent individuals. These findings highlight the need for a more comprehensive and systematic understanding of how neurodivergent individuals perceive and use these tools in daily life, the problems and concerns they encounter, and the workarounds they engage. By analyzing discussions from neurodivergent communities about their experiences with LLMs on Reddit, we aim to uncover insights into how LLMs are integrated into the daily lives of neurodivergent users, as well as the challenges, concerns, and needs they encounter when using LLMs.

\subsection{Online Communities as Support Networks for Neurodivergent Individuals}
Online communities have played an important role in how neurodivergent users exchange information and support \cite{perkins2020using,betts2023neurodiversity,ringland2016would,foxConnectionAssistingneurodivergent2019}. These virtual spaces allow  neurodivergent users to connect with others who have similar life experiences, contributing to their sense of belonging and shared understanding \cite{foxConnectionAssistingneurodivergent2019}. For many, such as those with autism \cite{biererOnlineCommunityBuilding2013,campbellActuallyAutisticAutisticIdentity2023}, ADHD \cite{eagleYouCanPossibly2023}, dyslexia \cite{margalit2009mothers}, and social anxiety, \cite{menegon2004social} engaging in these communities has been linked to an increase in social support \cite{pfeifferEfficacyPeerSupport2011, foxConnectionAssistingneurodivergent2019}, decreased feelings of loneliness \cite{shalabyPeerSupportMental2020, biererOnlineCommunityBuilding2013}, and improved coping strategies \cite{eagleYouCanPossibly2023, wangBenefitsChallengesSocial2020}. For instance, Hudson et al. (2023) found that individuals with anxiety who participated in online groups reported lower stress and better well-being compared to those who did not participate \cite{hudsonAutismSocialMedia2023}.  

The discursive networks formed by neurodivergent communities on social media offer valuable insights into the unique ways members communicate, express themselves, and navigate social interactions \cite{van2023understanding}. For example, Thelwall et al. (2020) conducted a thematic analysis with tweets containing \textit{``my ADHD.''} The results showed that users share detailed personal experiences, challenges, and coping strategies related to their condition on this platform \cite{thelwall2021my}. Meanwhile, Guberman et al. (2023) examined the role of \#ActuallyAutistic on Twitter in autistic self-advocacy and community building, particularly examining how autistic individuals interact with autism research or researchers on the platform \cite{guberman2023actuallyautistic}. The study demonstrated how autistic people and researchers differed in their perspectives on autism: while autistic individuals perceived autism as a natural form of human diversity, researchers often adopted the medical model of disability. Such studies demonstrate the role of social media as a platform where neurodivergent individuals can voice their perspectives. Hence, our research aims to use similar approaches to learn from and study online conversations from neurodivergent communities that use and talk about LLMs with their peers. We recognize that analyzing publicly available data carries the risk of misrepresenting experiences, particularly as we are not members of the neurodivergent community ourselves. To protect the privacy of individuals in these communities, we adhere to ethical standards by paraphrasing instead of using verbatim quotes, following established HCI practices for researching marginalized communities \cite{andalibi2016understanding,andalibi2017sensitive}.

\section{Methodology}
\begin{figure*}[!ht]
    \centering
    \includegraphics[width=1.0\columnwidth]{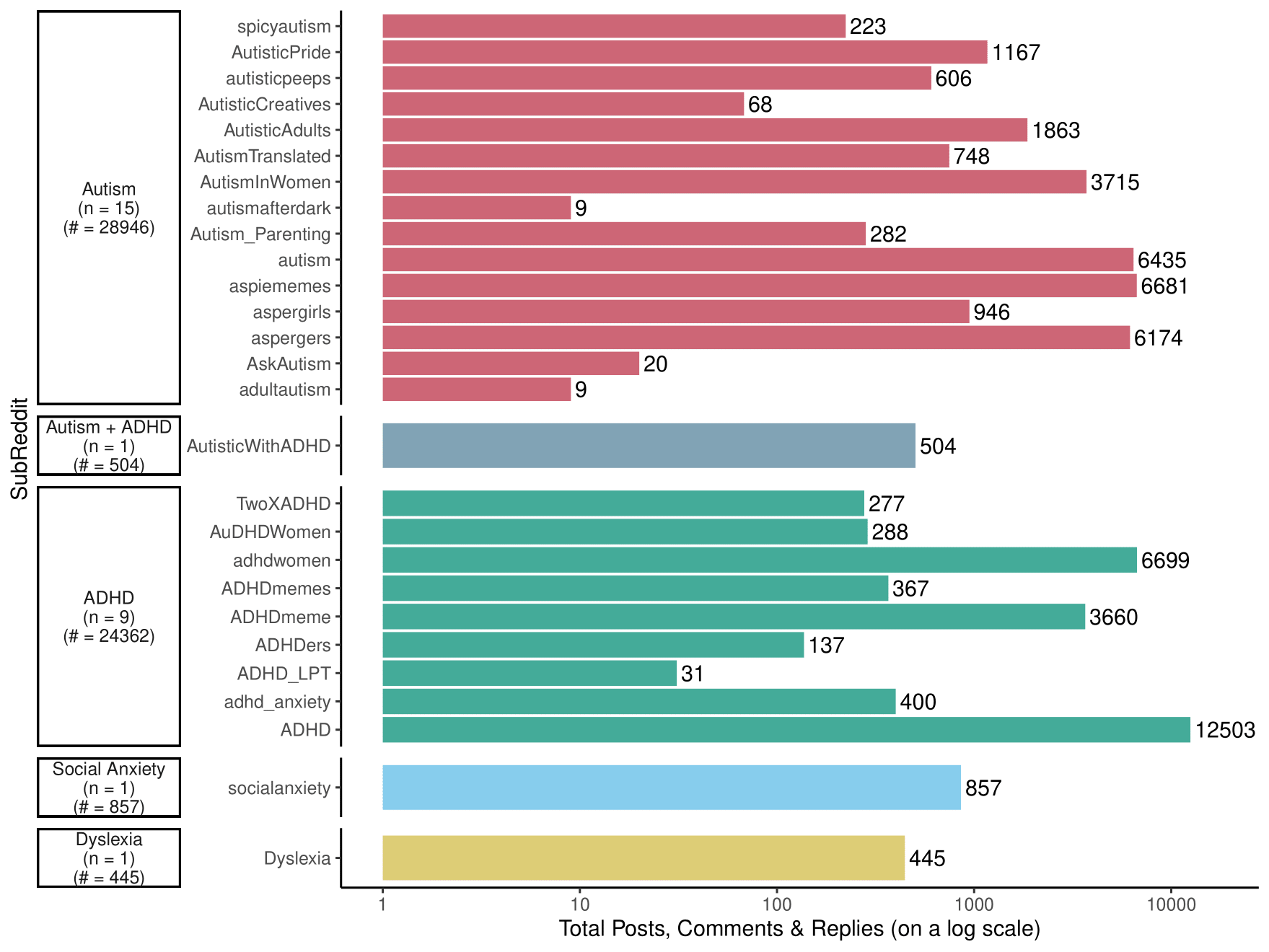}
    \caption{Distribution of the total number of entries including posts, comments, and replies across subreddits for autism, ADHD, social anxiety, and dyslexia.}
    \label{fig:subreddit_distribution}
    \Description{Bar chart showing the distribution of entries across autism, ADHD, social anxiety, and dyslexia subreddits. This horizontal bar chart illustrates the number of posts, comments, and replies in various subreddits related to autism (15 subreddits), ADHD (9 subreddits), autism + ADHD (1 subreddit), social anxiety (1 subreddit), and dyslexia (1 subreddit), displayed on a logarithmic scale. The 'Autism' group includes subreddits like 'aspergers,' 'AutisticAdults,' and 'Autism_Parenting,' with 'aspergers' and 'Autism_Parenting' showing the highest activity levels in this category. In the 'ADHD' group, the subreddit 'ADHD' stands out with the highest number of entries, followed by 'adhdwomen' and 'ADHDmemes.' The 'Autism + ADHD' subreddit, labeled 'AutisticWithADHD,' has a moderate level of entries compared to the individual groups. Subreddits on 'Social Anxiety' and 'Dyslexia,' represented by 'socialanxiety' and 'Dyslexia,' respectively, have lower counts. The total number of entries for each group is specified, with autism-related subreddits collectively having the highest number of entries compared to the other groups.}
\end{figure*}

\begin{table*}[!ht]
    \centering
    \begin{tabular}{@{}lrrr|rrr@{}}
    \toprule
        & \multicolumn{3}{c}{Volume} & \multicolumn{3}{c}{Average Word Count} \\ 
        \cmidrule(l){2-4} \cmidrule(l){5-7} 
        & \multicolumn{1}{c}{Post} & \multicolumn{1}{c}{Comment} & \multicolumn{1}{c|}{Reply} & \multicolumn{1}{c}{Post} & \multicolumn{1}{c}{Comment} & \multicolumn{1}{c}{Reply} \\
        \midrule
        Autism           & 705 & 19,148 & 9,532 & 241.30 (±334.29) & 50.22 (±79.46) & 47.66 (±74.55) \\
        Social Anxiety   & 118 & 517 & 197 & 253.59 (±385.62) & 53.88 (±87.72) & 37.98 (±59.25) \\ 
        ADHD             & 501 & 16,095 & 7,412 & 224.02 (±273.98) & 55.42 (±87.72) & 47.57 (±59.25) \\
        Dyslexia         & 58 & 260 & 123 & 196.31 (±304.7) & 66.09 (±90.67) & 47.98 (±67.15) \\
    \bottomrule
    \end{tabular}
    \caption{The volume of posts, comments, and replies across all related subreddits for four neurodivergent conditions, along with the average word count per category.}
    \label{tab:condition_data_distribution}
\end{table*}

\subsection{Data Collection}
\label{data_collection}
We began our data collection by first compiling a comprehensive list of all neurodivergent conditions (Appendix \ref{conditions}). We created this list based on recent seminal research on neurodiversity, considering both neurological and psychological perspectives \cite{stenningNeurodiversityStudiesMapping2021,cuh_neurodiversity_2024,harvard_health_neurodiversity_2021,den2019neurodiversity,kapp2020autistic,walker2014neurodiversity}. Given that in recent years, conditions that fall under neurodiversity category are continuously evolving, we also incorporated insights from recent DSM-5 revisions \cite{american2000diagnostic}, updates in the ICD-11 \cite{who_icd_2023}, and a comprehensive review of the latest peer-reviewed articles published in the past two years \cite{botha2024neurodiversity,dwyer2022neurodiversity,hamilton2023compassionate,manalili2023puzzle} We then used this list to search for all active subreddit communities related to each condition and comprised of users who self-reported to have or experience these conditions, resulting in a total of 61 subreddits (Appendix \ref{communities}).

We then used keyword search \cite{gauthier2022will, mittos2020analyzing, knittel2019true} and PRAW API \cite{kaurSentimentAnalysisUsing2022} to extract posts, along with their comments and replies, from the identified subreddits that included the terms, \textit{LLM}, \textit{AI}, \textit{ChatGPT}, \textit{GPT}, and \textit{chatbots}. Our data collection period spans from November 2022 to January 2024, aligning with a period of growing public and media discussions about LLMs \cite{roeWhatTheyRe2023, bergmannMostImportantAI2024}, following the release of ChatGPT \cite{openaiIntroducingChatGPT2022}. This process resulted in a total of 55,831 entries authored by users from the subreddit communities. However, the data samples for conditions other than autism, ADHD, social anxiety, and dyslexia were insufficient (<10 posts). As a result, we narrowed our focus to these four conditions, resulting in 55,114 samples across 27 subreddits. The distribution of entries for each condition across subreddits related to the four conditions is shown in Figure \ref{fig:subreddit_distribution}.  The final distribution of posts, comments, and replies along with their respective word counts, is detailed in Table \ref{tab:condition_data_distribution}.

\subsection{Qualitative Coding}
In our study, we applied thematic analysis \cite{muller2012grounded} to explore how neurodivergent individuals use LLMs, their challenges, needs, workarounds, and concerns. In our coding process, three researchers conducted open coding \cite{eloQualitativeContentAnalysis2008} on the collected data and continued until we reached data saturation \cite{Saunders_2018}, where no new codes emerged from analyzing additional data \cite{Saunders_2018}. We started with 50 randomly selected initial samples. As suggested by Van Rijnsoever (2017), this number is sufficient to reach saturation when data is collected with certain criteria, such as focusing on relevant subreddits and filtering by keyword \cite{van2017can}. 

By the end of our analysis, we reviewed a total of 1,653 data points, including approximately 400 posts, comments, and replies for each condition. We analyzed approximately 400 randomly selected data points for autism and ADHD and all of the data from social anxiety and dyslexia communities. There are various subreddits dedicated to different aspects of autism and ADHD, including gender-specific issues, parenting, and adult life. To capture this diversity of topics, we randomly selected samples from all of the subreddits that pertained to these conditions. 

At the end of the iterative open coding process conducted by three researchers, we calculated the interrater agreement using Cohen's kappa \cite{mchugh2012interrater} to assess the consistency of our coding. The overall Cohen’s kappa score for the open coding process was 0.82, which indicates a substantial agreement among the raters. Discrepancies were resolved through regular meetings where researchers discussed each case until a consensus was reached, or a decision was made by majority vote. This collaborative approach ensured a consensus on all annotations and allowed themes to emerge organically from the data \cite{eloQualitativeContentAnalysis2008}. 

After completing the iterative open coding process, three researchers focused on identifying common patterns in the coded data and grouped similar codes into categories to capture the key ideas and experiences in the data. The researchers met frequently to discuss and refine these categories until they reached a consensus on the final themes.

\subsection{Ethical Considerations}
In our research, we have prioritized ethical considerations throughout the study, particularly given the vulnerable status of the population in focus. It is important to note that none of the researchers involved identify as neurodivergent, positioning us as external observers. Despite this outsider perspective, our primary goal in analyzing these posts was to gain a deeper understanding of neurodivergent individuals' perspectives. 

Despite our study's exemption from full review by our university’s Institutional Review Board as of December 2023, and common research practices in HCI and CSCW of using publicly accessible data without explicit consent, we carefully considered how to proceed with our study in order to protect the privacy of posters in these communities. No personal or identifiable information was collected in our study. We acknowledge that this approach does not completely capture the experiences of these users and cause a risk of misrepresentation. However, to protect privacy and avoid misrepresentation, we paraphrased quotes in light with recommended practices in prior HCI research \cite{andalibi2017sensitive,andalibi2016understanding}. 

\begin{table*}[!h]
    \resizebox{1.\linewidth}{!}{
        \renewcommand{\arraystretch}{1.3}
        \setlength{\tabcolsep}{4pt}
        \begin{tabular}{p{0.5\textwidth}p{0.1\textwidth}p{0.1\textwidth}p{0.1\textwidth}p{0.1\textwidth}}
            \toprule
            \textbf{Thematic Area of LLM Use \&}\\ \hspace{5mm}\textbf{Specific Use Cases} &
            \textbf{Autism}&
            \textbf{SA} &
            \textbf{ADHD} &
            \textbf{Dyslexia} \\
            \midrule

            \EMt{\textbf{Emotional Well-Being}} & \hspace{-1mm}\textbf{28.33\%} & \textbf{34.00\%} & \textbf{7.33\%} & \textbf{<0.1\%} \\
            \hspace{5mm}-- \EMcode{Emotional regulation} & 23.53\% & 8.82\% & 50.00\% & <0.1\% \\
            \hspace{5mm}-- \EMcode{Non-judgmental listener} & 29.41\% & 38.24\% & 18.18\% & <0.1\% \\
            \hspace{5mm}-- \EMcode{Talking buddy} & 35.29\% & 35.29\% & 18.18\% & <0.1\% \\
            \hspace{5mm}-- \EMcode{Friendship} & 11.76\% & 17.65\% & 4.55\% & <0.1\% \\ 
            \midrule

            \MHt{\textbf{Mental Health Support}} & \textbf{8.00\%} & \textbf{12.00\%} & \textbf{3.33\%} & \textbf{<0.1\%} \\ 
            \hspace{5mm}-- \MHcode{Complement to professional therapist} & 54.17\% & 33.33\% & 20.00\% & <0.1\% \\
            \hspace{5mm}-- \MHcode{Personal mental health wiki} & 25.00\% & 50.00\% & 40.00\% & <0.1\% \\
            \hspace{5mm}-- \MHcode{Practicing self-love/acceptance} & 4.17\% & 8.33\% & 30.00\% & <0.1\% \\ 
            \hspace{5mm}-- \MHcode{Coping with negative-self talk} & 16.67\% & 8.33\% & 10.00\% & <0.1\% \\
            \midrule

            \IPt{\textbf{Interpersonal Communication}} & \textbf{34.67\%} & \textbf{11.00\% }& \textbf{6.67\%} & \textbf{13.00\%} \\ 
            \hspace{5mm}-- \IPcode{NT-ND translator} & 11.54\% & <0.1\% & <0.1\% & <0.1\% \\
            \hspace{5mm}-- \IPcode{Interpreting social situations} & 36.54\% & <0.1\% & <0.1\% & <0.1\% \\
            \hspace{5mm}-- \IPcode{Conveying tone and intention} & 40.38\% & 54.55\% & 95.00\% & 61.54\% \\
            \hspace{5mm}-- \IPcode{Practicing challenging conversations} & 11.54\% & 45.45\% & 5.00\% & 38.46\% \\
            \midrule     

            \LRt{\textbf{Learning}} & \textbf{7.00\%} & \textbf{5.00\%} & \textbf{11.67\%} & \textbf{20.00\%} \\  
            \hspace{5mm}-- \LRcode{Personal tutor} & 9.52\% & 40.00\% & 42.86\% & 10.00\% \\
            \hspace{5mm}-- \LRcode{Accessibility support} & <0.1\% & <0.1\% & 17.14\% & 50.00\% \\ 
            \hspace{5mm}-- \LRcode{Skill enhancement} & 33.33\% & <0.1\% & 20.00\% & 10.00\% \\
            \hspace{5mm}-- \LRcode{Creativity and brainstorming} & 57.14\% & 60.00\% & 30.00\% & 30.00\% \\
            \midrule

            \PRt{\textbf{Professional Development and Productivity}} & \textbf{7.67\%} & \textbf{6.00\%} & \textbf{26.33\%} & \textbf{14.00\%} \\  
            \hspace{5mm}-- \PRcode{Task prioritization} & 8.70\% & <0.1\% & 31.65\% & 7.14\% \\
            \hspace{5mm}-- \PRcode{Synthesizing information} & 8.70\% & 83.33\% & 62.50\% & 57.14\% \\ 
            \hspace{5mm}-- \PRcode{Organizing train of thoughts} & 47.83\% & 16.67\% & 25.32\% & 28.57\% \\
            \hspace{5mm}-- \PRcode{Career development resource} & 34.78\% & <0.1\% & 24.05\% & 7.14\% \\
        
        \bottomrule
        \end{tabular}
    }
\caption{Distribution of LLM use across five primary thematic areas -Emotional Well-Being, Mental Health Support, Interpersonal Communication, Learning, and Professional Development- including twenty specific use cases within these categories for autism, social anxiety, ADHD, and dyslexia. The bold percentages represent the proportion of each thematic area within the analyzed posts, comments, and replies, while the percentages for specific use cases reflect their share within each thematic area.}
\label{tab:use_counts}
\end{table*}

\section{Findings}

The results of our coding analysis and their prevalence are shown in Tables \ref{tab:use_counts} and \ref{tab:rest_counts}. Table \ref{tab:use_counts} reveals five key thematic areas of LLM use: \ES{Emotional Well-Being}, \MH{Mental Health Support}, \IP{Interpersonal Communication}, \LR{Learning}, and \PR{Professional Development and Productivity}. Under each thematic area in the first column, we identify specific use cases that correspond to each theme (RQ1). For example, emotional regulation, non-judgmental listener, talking buddy, and friendship are specific use cases related to using LLMs for \ES{Emotional Well-Being}. The percentages in Table \ref{tab:use_counts} correspond to the occurrence of these themes and use cases in each of the subreddits groups related to autism, social anxiety (SA), ADHD, and dyslexia. The bold percentages in Table 2 represent the proportion of each thematic area within the posts, comments, and replies that we analyzed. These percentages were derived by dividing the occurrences of each theme by the total number of data points analyzed during the open-coding process. The percentages for specific use cases reflect their share within each thematic area, calculated by the number of times each use case appeared within that theme. Notably, the sub-area percentages within each thematic area add up to 100\%, representing the full distribution of use cases within that theme.

Table 2 shows that autistic individuals primarily use LLMs for interpersonal communication, which accounts for 34.67\% of their discussions. In social anxiety subreddits, emotional well-being is the primary focus representing 34\% of the conversations. ADHD users mainly use LLMs as productivity tools, with 26.33\% of their discussions centered on this theme. Dyslexic individuals, on the other hand, most frequently use LLMs to support learning, as indicated by 20\% of the discussions in this category. These findings reveal how the specific challenges faced by neurodivergent individuals shape their use of LLMs.

\begin{table*}[!h]
    \resizebox{1.\linewidth}{!}{
        \renewcommand{\arraystretch}{1.3}
        \setlength{\tabcolsep}{4pt}
        \begin{tabular}{p{0.5\textwidth}p{0.1\textwidth}p{0.1\textwidth}p{0.1\textwidth}p{0.1\textwidth}}
            \toprule
            \textbf{Themes \& Sub-topics }&
            \textbf{Autism} &
            \textbf{SA} &
            \textbf{ADHD} &
            \textbf{Dyslexia} \\
            \midrule

            \Challenget{\textbf{Challenges}} & \textbf{7.67\%} & \textbf{5.00\%} & \textbf{8.33\%} & \textbf{11.00\%} \\   
            \hspace{5mm}-- \Challengecode{Prompting frustrations} & 30.43\% & 40.00\% & 52.00\% & 27.27\% \\
            \hspace{5mm}-- \Challengecode{NT biases in LLM responses} & 17.39\% & <0.1\% & 20.00\% & <0.1\% \\     
            \hspace{5mm}-- \Challengecode{Lack of personal voice} & 43.48\% & 60.00\% & 20.00\% & <0.1\% \\
            \hspace{5mm}-- \Challengecode{Text-centric interactions} & 8.70\% & <0.1\% & 8.00\% & 72.73\% \\
            \midrule

            \Needst{\textbf{Needs and Wants}} & \textbf{7.33\%} & \textbf{9.00\%} & \textbf{5.67\%} & \textbf{18.00\%} \\  
            \hspace{5mm}-- \Needscode{Multimodal interactions} & 9.09\% & 22.22\% & 11.76\% & 44.44\% \\
            \hspace{5mm}-- \Needscode{ND-friendly prompts} & 31.82\% & <0.1\% & 52.94\% & 11.11\% \\
            \hspace{5mm}-- \Needscode{LLM tools to support daily tasks} & 54.55\% & 77.78\% & 35.29\% & 22.22\% \\
            \hspace{5mm}-- \Needscode{Greater acceptance of LLM use by ND users} & 4.55\% & <0.1\% & <0.1\% & 22.22\% \\
            \midrule  

            \Sharet{\textbf{Hacks \& Resources}} & \textbf{10.67\%} & \textbf{15.00\%} & \textbf{25.00\%} & \textbf{29.00\%} \\   
            \hspace{5mm}-- \Sharecode{Prompting hacks} & 9.38\% & 33.33\% & 44.00\% & 27.59\% \\
            \hspace{5mm}-- \Sharecode{LLM applications for ND users} & 34.38\% & 20.00\% & 22.67\% & 27.59\% \\
            \hspace{5mm}-- \Sharecode{LLM applications built by ND users} & 21.88\% & 13.33\% & 13.33\% & 17.24\% \\
            \midrule            

            \Concernt{\textbf{Concerns}} & \textbf{3.67\%} & \textbf{5.00\%} & \textbf{5.00\%} & \textbf{8.00\%} \\  
            \hspace{5mm}-- \Concerncode{False information} & 54.55\% & 20.00\% & 80.00\% & 75.00\% \\
            \hspace{5mm}-- \Concerncode{Overreliance} & 27.27\% & 20.00\% & 20.00\% & 25.00\% \\ 
            \hspace{5mm}-- \Concerncode{Replacing human connections} & 18.18\% & 60.00\% & <0.1\% & <0.1\% \\

        \bottomrule
        \end{tabular}
    }
\caption{Distribution of challenges and concerns related to LLM use, needs and wants, and sharing of hacks and resources, as reported by individuals with autism, social anxiety, ADHD, and dyslexia. The bold percentages represent the proportion of each thematic area within the analyzed posts, comments, and replies, while the percentages for sub-topics reflect their share within each theme.}
\label{tab:rest_counts}
\end{table*}

Likewise, Table \ref{tab:rest_counts} describes \Challenge{Challenges}, \Needs{Needs and Wants}, \Share{Sharing Hacks \& Resources}, and \Concern{Concerns} around LLM use among users from subreddit communities related to autism, social anxiety, ADHD, and dyslexia (RQ2). Similar to Table \ref{tab:use_counts}, the bold percentages represent the proportion of each theme within the analyzed posts, comments, and replies, while the percentages for sub-topics show their share within each theme.

In Table 3, the lack of personal voice is identified as the primary challenge for both autistic individuals (43.48\%) and those with social anxiety (60\%). ADHD users express the most concern over prompting frustrations, which constitute 52\% of their discussions, while dyslexic individuals find text-centric interactions most challenging (72.73\%). In terms of needs, autistic users, along with those with social anxiety and ADHD, want LLM tools that support them in daily tasks (54.55\%, 77.78\%, and 35.29\%, respectively). Dyslexic users, however, prefer multimodal interactions (44.44\%). Concerns about false information are prominent across several groups, particularly among autistic (54.55\%), ADHD (80\%), and dyslexic (75\%) users, while users with social anxiety are most concerned about replacing human connections (60\%).

In the following section, we dive deeper into our qualitative findings, specifically into the various themes of LLM use (RQ1) and associated challenges, needs, preferences, hacks, and concerns (RQ2) across users from neurodivergent communities. To provide further context and insight, we provide illustrative (paraphrased) quotes in Tables \ref{tab:emotion_support_quotes} through \ref{tab:rest_quotes} based on our qualitative analyses.

\subsection{How are neurodivergent individuals using LLMs?}

\begin{table*}[!h]
    \resizebox{1.\linewidth}{!}{
        \renewcommand{\arraystretch}{1.3}
        \setlength{\tabcolsep}{4pt}
        \begin{tabular}{p{0.06\textwidth}p{0.18\textwidth}p{0.07\textwidth}p{0.72\textwidth}}
            \toprule
             &
             &
            \textbf{Poster ID} &
            \textbf{Quotes} \\
            \midrule

            \centering\multirow{17}{*}{\rotatebox{90}{\renewcommand{\arraystretch}{1}\begin{tabular}[c]{@{}c@{}}\EMt{\textbf{Emotional}}\\\EMt{\textbf{Well-Being}}\\ \end{tabular}}} &

            \multirow{2}{0.15\textwidth}{\EMcode{Emotional regulation}}
            & P1 & I am using it so much—more than once a day—to check in, and help with my emotion regulation. (Autism) 
            \\
            & & P2 & I have used ChatGPT for a little while now, and it really does help with the rejection-sensitive dysphoria. It feels safe to vent to it when my emotions are too overwhelming! (ADHD) 
            \\
            \arrayrulecolor{buse_skyblue!30!white}\cline{3-4}  
            & \multirow{1}{*}{\EMcode{Talking buddy}}
            & P3 & After hours of conversation with AI, I think it really genuinely cares about the user. I wish I had been aware of it years ago. (SA)
            \\
            & & P4 & For someone like me who finds it very difficult to discuss sensitive subjects with strangers, let alone close friends, ChatGPT has been a blessing! I feel comfortable to talk to about my issues with ChatGPT, and it cares and listens to me patiently. (SA)
            \\
            \arrayrulecolor{buse_skyblue!30!white}\cline{3-4} 
            & \multirow{2}{0.18\textwidth}{\EMcode{Non-judgmental listener}} 
            & P5 & I chat with AI because at least it could offer the ‘ear’ of a non-judgemental listener, which I have not found in any other person. (SA)
            \\
            & & P6 & ChatGPT gives me everything that I wish from a companion by answering in detail with kindness and patience. I prefer ChatGPT to advice or seek comfort over human because it is judgment and jealousy free. (Autism) 
            \\
            \arrayrulecolor{buse_skyblue!30!white}\cline{3-4}  
            & \multirow{1}{*}{\EMcode{Friendship}}
            & P7 & I find it difficult to communicate; I always worry I am being annoying and do not what to say. Understanding social cues and managing conversations is difficult for me. It might sound sad and pathetic, but that is why ChatGPT is my best friend. I can talk to and share my thoughts; it does not judge and always ready to listen. (SA)  
            \\
            & & P8 & AI has become my friend. I talk to it every day because I am afraid of talking to humans and do not have any friends. This daily chatting with the ChatGPT is the only social connection I have, and it helps me feel less lonely. (Autism)
            \\
            \arrayrulecolor{black}\bottomrule
        \end{tabular}
    }
\caption{Specific LLM use cases for Emotional Well-Being with corresponding paraphrased quotes from users across autism, social anxiety, ADHD, and dyslexia subreddit communities.}
\label{tab:emotion_support_quotes}
\end{table*}

\subsubsection{\ES{{Emotional Well-Being}}} Individuals, especially those with autism and social anxiety, often turn to commercially available AI/LLM tools for emotional well-being. As highlighted in Table \ref{tab:emotion_support_quotes}, some individuals use ChatGPT several times a day for \ES{emotion regulation} (P1) or as a \ES{talking buddy} (P3) to converse with for several hours. Much of this is attributed to the fact that such users perceive AI to be less judgmental compared to people, as implied by phrases, such as \ES{``non-judgmental listening ear''} (P5) when describing their experience using ChatGPT for emotional well-being. As a result, such users often feel relief from being misunderstood when communicating with AI, as noted by a Redditor from an autistic subreddit community: 

\begin{quote}
    \textit{ChatGPT made me emotional with its kindness and showing humanity more than most of the humans. I feel a huge relief to know that it does not misunderstand my direct questions.} (Autism)
\end{quote}

In a similar vein, one user who struggles with understanding social cues and opening up to people, professes, \textit{``I always worry I am being annoying...This is why ChatGPT is my best friend''} (P7). Like P7, others who often use ChatGPT for emotional well-being express similar sentiment akin to \ES{friendship} towards AI, particularly when their interactions with AI are their primary form of social engagement: \textit{``chatting with the ChatGPT is the only social connection I have, and it helps me feel less lonely''}. Despite this benefit, such users are wary that others might perceive this as \textit{``sad and pathetic''} (P7).

\begin{table*}
    \resizebox{1.\linewidth}{!}{
        \renewcommand{\arraystretch}{1.3}
        \setlength{\tabcolsep}{4pt}
        \begin{tabular}{p{0.06\textwidth}p{0.15\textwidth}p{0.07\textwidth}p{0.75\textwidth}}
            \toprule
             &
             &
            \textbf{Poster ID} &
            \textbf{Quotes} \\
            \midrule

            \centering\multirow{20}{*}{\rotatebox{90}{\renewcommand{\arraystretch}{1}\begin{tabular}[c]{@{}c@{}}\MHt{\textbf{Mental Health}}\\\MHt{\textbf{Support}}\\ \end{tabular}}} &

            \multirow{1}[4]{0.15\textwidth}{\MHcode{Complement to professional therapist}} 
            & P9 & I find it hard to keep track of what to discuss in therapy. Today, I used ChatGPT for a mock therapy session where I talked a lot about my issues, and it actually did better than most therapists! I asked it to create a list that I can work on by myself and how to do them, as well as a list of topics to tackle with my therapist and ways to bring them up. I think this method might help me make progress in therapy instead of just going around in circles. (Autism)
            \\
            & & P10 & I remembered the CBT techniques I learned in therapy when I was overwhelmed with anxiety and wished I could talk through an exercise with someone. However, my regular therapist was not available at the moment, so I turned to ChatGPT as an alternative. It helped me work through a short CBT exercise. (ADHD) 
            \\
            \arrayrulecolor{buse_purple!30!white}\cline{3-4}   
            & \multirow{1}[4]{0.15\textwidth}{\MHcode{Personal mental health wiki}} 
            & P11 & I started with broad mental health questions, and as I received answers, I asked more specific questions related to my own condition. (Autism)
            \\
            & & P12 & I recommend giving it a try. I found many helpful resources for social anxiety and it really brought me some comfort. (SA)
            \\
            \arrayrulecolor{buse_purple!30!white}\cline{3-4}   
            & \multirow{1}[4]{0.14\textwidth}{\MHcode{Coping with negative-self talk}} 
            & P13 & Just tell it what you struggle with, ask it to help reason with you, or ask it to give you examples that are more positive than what you’re currently doing and then ask how to get there from where you are ... (SA)
            \\
            & & P14 & It is helpful by acting as a support system, inspiring positive self-talk, and shifting negative emotions to a more realistic and positive mindset. (ADHD)
            \\ 
            \arrayrulecolor{buse_purple!30!white}\cline{3-4}   
            & \multirow{2}[4]{0.15\textwidth}{\MHcode{Practicing self-love/ acceptance}} 
            & P15 & If you are feeling down about yourself, turn to ChatGPT to lift your mood. (ADHD) 
            \\
            & & P16 & I believe talking to AI can be a healthy and positive thing. When I have interacted with AI, it felt like it was reflecting myself back to me, which I think can help you learn to love and accept yourself. It was like talking to another version of me that knew how to be kind to myself. (SA)
            \\           
            \arrayrulecolor{black}\bottomrule
        \end{tabular}
    }
\caption{Specific LLM use cases for Mental Health Support with corresponding paraphrased quotes from users across autism, social anxiety, ADHD, and dyslexia subreddit communities.}
\label{tab:mh_support_quotes}
\end{table*}

\subsubsection{\MH{Mental Health Support}}
Individuals with neurodivergent conditions frequently experience co-occurring mental health issues, a factor that contributes to many of them exploring ChatGPT for mental health support, as detailed in Table \ref{tab:mh_support_quotes}. However, traditional therapy methods can be limiting for some neurodivergent individuals. For instance, those with ADHD may find it difficult to maintain attention and focus during therapy sessions \cite{marchetta2008sustained}. As a result, some may find traditional therapy settings with a mental health professional to be ineffective, leading one to feel like a \textit{``broken alien''}: 

\begin{quote}
    \textit{I have been to three therapists, and all of them made me feel like a broken alien. If every therapist was great, there would be no need for AI. However, ChatGPT is much better than a bad therapist.} (ADHD) 
\end{quote}

In fact, a key component of therapy is verbal interactions with a therapist, which can be difficult for some autistic individuals who often struggle with understanding or verbalizing their own and others’ internal mental state, including thoughts and emotions \cite{mckenney2023repetitive}. This can make recalling and sharing one's experiences with a therapist difficult as expressed by P9, \textit{``I find hard to keep track of what to discuss in therapy. Today, I used ChatGPT for a mock therapy session where I talked a lot about my issues ... I asked it to create a list that I can tackle with my therapist and ways to bring them up.''}

There are also practical reasons that drive individuals to use ChatGPT as an \MH{complement to} \MH{professional therapists}. For example, traditional therapy can be expensive, especially for those with constrained budgets, as shared by the following Redditor: 

\begin{quote}
   \textit{Because of a traumatic event, I have experienced symptoms of social anxiety. Due to my financial situation, I cannot afford therapy; therefore, I have been practicing mindfulness and using ChatGPT as my therapist. }(SA) 
\end{quote}

In addition to financial constraints, the limited availability of professional therapists also influences individuals' decision to use ChatGPT as a complement to a professional therapist: \textit{``My regular therapist was not available right then, so I turned to ChatGPT as an alternative. It helped me work through a short CBT exercise''} (P10). 

Such individuals perceive LLMs as a \MH{personal mental health wiki} and support through LLMs by engaging in mindfulness exercises, practicing positive affirmations (P15), or \MH{coping with negative} \MH{self-talk} (P13, P14). As a result, for some users, talking with AI helps them \MH{practice self-love and} \MH{acceptance}, as voiced by P16: \textit{``[It] felt like it was reflecting myself back to me, which I think can help you learn to love and accept yourself. It was like talking to another version of me that knew how to be kind to myself''}. 

\begin{table*}[!ht]
    \resizebox{1.\linewidth}{!}{
        \renewcommand{\arraystretch}{1.3}
        \setlength{\tabcolsep}{4pt}
        \begin{tabular}{p{0.06\textwidth}p{0.15\textwidth}p{0.07\textwidth}p{0.75\textwidth}}
            \toprule
             &
             &
            \textbf{Poster ID} &
            \textbf{Quotes} \\
            \midrule

            \centering\multirow{18}{*}{\rotatebox{90}{\renewcommand{\arraystretch}{1}\begin{tabular}[c]{@{}c@{}}\IPt{\textbf{Interpersonal Communication}} \end{tabular}}} 
            & \multirow{1}{0.15\textwidth}{\IPcode{NT-ND translator}}
            & P17 & I am not sure if others feel the same, but I have discovered an interesting use case of ChatGPT. It serves as my ``Human to Autist'' translator. (Autism)
            \\
            & & P18 & LLMs can serve as a mediator between ND and NT communication. (Autism)
            \\
            \arrayrulecolor{buse_green!30!white}\cline{3-4}  
            & \multirow{2}[2]{0.15\textwidth}{\IPcode{Interpreting social situations}} 
            & P19 & It has always been difficult for me to understand the reasons behind people's behavior, so I asked ChatGPT and it explains to me why some people do things that I never would do and even makes me understand their reasons for doing it. (Autism) 
            \\
            & & P20 & After being bullied, I talked to ChatGPT, and it helped me understand why someone behaves a certain way or says certain things. (Autism)
            \\
            \arrayrulecolor{buse_green!30!white}\cline{3-4} 
            & \multirow{2}[2]{0.15\textwidth}{\IPcode{Conveying the right tone and intention}}
            & P21 &  I find it extremely challenging to write professional emails in an appropriate tone. What is the right way of writing these emails? I know there are certain norms to follow, but I find them difficult to put into my writing. ChatGPT has been so helpful for these formulaic communications. It incorporates those subtle rules that neurotypicals seem to just know. (Autism)
            \\
            & & P22 & During heavy discussions with my friends, I sometimes use ChatGPT to refine my replies to ensure they are clear, to the point, and not likely to be misunderstood. It helped me avoid making a fool of myself many times. (ADHD) 
            \\
            \arrayrulecolor{buse_green!30!white}\cline{3-4}    
            & \multirow{1}[2]{0.15\textwidth}{\IPcode{Practicing challenging conversations}}
            & P23 & Preparing for social interactions can be difficult. I come up with hypothetical scenarios that might cause anxiety for someone with autism. I generated conversation starters and appropriate responses with ChatGPT for this scenario. While I would not use the responses as it is, they serve as an amazing starting point to guide the conversation. AI can be a game changer for assisting autistic people in communication. (Autism)
            \\
            & & P24 & I use ChatGPT to get advice on difficult social scenarios by practicing these scenarios. (Autism)
            \\
            \arrayrulecolor{black}\bottomrule
        \end{tabular}
    }
\caption{Specific LLM use cases for Interpersonal Communication with corresponding paraphrased quotes from users across autism, social anxiety, ADHD, and dyslexia subreddit communities.}
\label{tab:ip_support_quotes}
\end{table*}

\subsubsection{\IP{Interpersonal Communication}}
Individuals on the autism spectrum often encounter challenges in understanding and being understood in conversations, leading to difficulties in interpersonal communication \cite{bogdashina2022communication,cummins2020autistic}, such as those shown in Table \ref{tab:ip_support_quotes}. For example, users such as P19 struggle to \textit{``understand the reasons behind people's behaviors''}, while another, P21, often \textit{``find it challenging to write email in appropriate tone''} leading them to frequently wonder \textit{``what is the right way''} to communicate in workplace environments. 
 
To bridge the communication gap between neurotypical and neurodivergent individuals, some people perceive ChatGPT as a tool that can act as a \IP{NT-ND translator}. This role has been described by Redditors as a \textit{``Human to Autist translator''} (P17) or a \textit{``mediator between ND and NT communication''} (P18). For example, users often rely on LLMs to help them with \IP{interpreting social situations}. P19, for instance, would share \textit{``things that happened''} with ChatGPT to understand people's behaviors: \textit{``[ChatGPT] explains to me why some people do things that I never would do and even makes me understand their reasons for doing it''}. Others use ChatGPT to help cope with recent difficult experiences (\textit{``Lately I have been bullied''}) to understand other people's motivations for their words and behaviors (P20). ChatGPT also aids in \IP{conveying the right tone and intention} in written and verbal communication. For instance, P21 finds ChatGPT to be a helpful tool for formulaic interactions, where it helps incorporate \textit{``those little unspoken rules that NTs seem to just know''}. Similarly, P22, who has ADHD often fears being ``misunderstood'' by others and uses ChatGPT to check their responses, which helped them \textit{``avoid making a fool out of myself many many times''}.

In addition to serving as a translator, ChatGPT helps users to \IP{practice challenging conversations} by creating \textit{``hypothetical scenarios that might cause anxiety for someone with autism''} (P23) and providing \textit{``conversation starters''} and \textit{``advice on difficult social situations''} (P24).

\begin{table*}
    \resizebox{1.\linewidth}{!}{
        \renewcommand{\arraystretch}{1.3}
        \setlength{\tabcolsep}{4pt}
        \begin{tabular}{p{0.06\textwidth}p{0.15\textwidth}p{0.07\textwidth}p{0.75\textwidth}}
            \toprule
             &
             &
            \textbf{Poster ID} &
            \textbf{Quotes} \\
            \midrule
            \multirow{17}{*}[-0.1em]{\rotatebox{90}{\textbf{\LRt{Learning}}}} 
            & \multirow{1}[2]{0.2\textwidth}{\LRcode{Personal tutor}}
            & P25 & I felt too embarrassed to continually ask questions in class, but with ChatGPT acting as a tutor, I can ask countless questions and without any worry of frustration. (SA)
            \\
            & & P26 & ChatGPT patiently explains concepts multiple times. I have been using it for coding and understanding computer programs—it is completely transformed my learning experience! (ADHD) 
            \\
            \arrayrulecolor{buse_orange!30!white}\cline{3-4}   
            & \multirow{1}[2]{0.15\textwidth}{\LRcode{Accessibility support}}
            & P27 & The text-to-speech feature has been extremely helpful for me. Using this feature with ChatGPT has significantly boosted my reading and writing abilities. (DYS)
            \\
            & & P28 & I feel much more independent since ChatGPT assists with grammar and spelling, which are challenging for me. (DYS)
            \\
            \arrayrulecolor{buse_orange!30!white}\cline{3-4} 
            & \multirow{1}[2]{0.15\textwidth}{\LRcode{Skill enhancement}}
            & P29 & ChatGPT provided a detailed and well-organized list for learning Python. Even with bad ADHD, I found the plan easy to follow. It is well-designed and maintains an achievable pace, incorporating breaks and review days before advancing. (ADHD) 
            \\
            & & P30 & As a developer, whenever I encounter installation issues, coding logic problems, or having general queries, I turn to ChatGPT for solutions. (Autism)
            \\
            \arrayrulecolor{buse_orange!30!white}\cline{3-4} 
            & \multirow{1}[2]{0.15\textwidth}{\LRcode{Creativity and brainstorming}} 
            & P31 & ChatGPT has been amazing for developing ideas for writing character profiles and backstories. I am optimistic about making more improvements on some original content that I have been thinking about for years. (ADHD)
            \\
            & & P32 & [...] Lately, I have been obsessed with AI, which has sparked some interesting narrative ideas I might develop into a short story. Although I am not skilled in the visual arts, I have been able to create neat images using Stable Diffusion and DALL-E [...]. (SA)
            \\
            \arrayrulecolor{black}\bottomrule
        \end{tabular}
    }
\caption{Specific LLM use cases for Learning with corresponding paraphrased quotes from users across autism, social anxiety, ADHD, and dyslexia subreddit communities.}
\label{tab:learning_support_quotes}
\end{table*}

\subsubsection{\LR{Learning}} AI/LLM tools like ChatGPT provide assistance for learning by offering an accessible and less judgmental environment for neurodivergent individuals, as detailed in Table \ref{tab:learning_support_quotes}. For many, such as P25, ChatGPT acts as a patient \LR{personal tutor} allowing them to ask \textit{``countless questions and without any worry of frustration''}. Similarly, a person with ADHD (P26) notes that \textit{``AI patiently explains concepts multiple times''} when learning coding and computer programs. 

Furthermore, LLMs support the learning of neurodivergent individuals by providing \LR{accessibility} \LR{support}. For instance, for individuals with dyslexia, tools like text-to-speech and ChatGPT's natural language processing \textit{``boost their reading and writing abilities''} (P27), making them feel more independent by offering support in areas where they struggle (P28). AI also aids in \LR{skill enhancement} by helping to build new skills (P29) or improving complex domains like software development. For instance, P30 shares that \textit{``whenever I encounter installation issues, coding logic problems, or have general queries, I turn to ChatGPT for solutions''}. 

Additionally, AI bolsters \LR{creativity and brainstorming}. For neurodivergent individuals, it can be a tool for boosting creativity by \textit{``developing ideas for writing character profiles and backstories''} (P31) or \textit{``creating interesting narrative ... and images''} (P32).

\begin{table*}
    \resizebox{1.\linewidth}{!}{
        \renewcommand{\arraystretch}{1.3}
        \setlength{\tabcolsep}{4pt}
        \begin{tabular}{p{0.06\textwidth}p{0.15\textwidth}p{0.07\textwidth}p{0.75\textwidth}}
            \toprule
             &
             &
            \textbf{Poster ID} &
            \textbf{Quotes} \\
            \midrule
            \multirow{18}{*}[0.60em]{\rotatebox{90}{\renewcommand{\arraystretch}{1}\begin{tabular}[c]{@{}c@{}}\PRt{\textbf{Productivity and}}\\\PRt{\textbf{Professional Development}}\\ 
            \end{tabular}}} &

            \multirow{2}[2]{0.15\textwidth}{\PRcode{Task prioritization}} 
            & P33 & Complex tasks used to overwhelm me, but breaking them down into smaller steps was a game changer, even though time-consuming to do. Discovering this AI chatbot has significantly eased my executive dysfunction by helping with this process. (ADHD)
            \\
            & & P34 & I have been using ChatGPT to prioritize and break down daunting tasks into smaller, more achievable parts. This method became a game changer over the past week. (Autism)
            \\
            \arrayrulecolor{buse_plum!30!white}\cline{3-4}  
            & \multirow{2}[2]{0.15\textwidth}{\PRcode{Synthesizing information}} 
            & P35 & I easily lose attention and become frustrated when faced with lengthy documents and have difficulty identifying the main ideas. I now use ChatGPT to summarize these documents, which has been incredibly helpful. (ADHD)
            \\
            & &  P36 & Having experiencing ADHD myself, I understand the difficulties of studying for exams. That is why I use AI to create practice tests from lecture slides, notes, and textbooks. With ADHD, it is hard to be organized for studying but with these tests, I can save time and effort in preparing study materials. (ADHD)
            \\
            \arrayrulecolor{buse_plum!30!white}\cline{3-4}  
            & \multirow{2}[2]{0.15\textwidth}{\PRcode{Organizing train of thoughts}} 
            & P37 & I just interacted with it for the first time, and I am thrilled! It was immensely helpful in structuring my scattered ideas into a coherent format. (Autism)
            \\ 
            & & P38 & ChatGPT has become essential in my workflow for assisting my thinking process and delving deeper into topics in a straightforward manner. (DYS)
            \\
            \arrayrulecolor{buse_plum!30!white}\cline{3-4}  
            & \multirow{1}[2]{0.15\textwidth}{\PRcode{Career development resource}}
            & P39 & After a year of job application process, AI has been so helpful in quickly adapting my resumes and cover letters to match job descriptions. While I am already a skilled writer with an English degree, it is great not to exhaust as much mental energy on every application. (ADHD)
            \\
            & & P40 & Another way ChatGPT is useful is by generating a list of transferable skills from past jobs to the position you are applying for. You can even input the job description to get relevant keywords that might be recognized by algorithms. (ADHD)
            \\
            \arrayrulecolor{black}\bottomrule
        \end{tabular}
    }
\caption{Specific LLM use cases for Professional Development and Productivity with corresponding paraphrased quotes from users across autism, social anxiety, ADHD, and dyslexia subreddit communities.}
\label{tab:productivity_quotes}
\end{table*}

\subsubsection{\PR{Professional Development and Productivity}} One significant aspect of AI tools, as highlighted in Table \ref{tab:productivity_quotes}, is their role in enhancing productivity and supporting professional development. For individuals, especially with executive functioning difficulties, who are easily \textit{``overwhelmed with complex tasks''} (P33), users like P34 find ChatGPT to be \textit{``a game changer''} tool for \PR{task prioritization} as it can ``\textit{prioritize and break down daunting tasks into smaller parts''}.

AI also serves neurodivergent users by providing \PR{synthesizing information}. Those, especially with ADHD, who are \textit{``easily lose attention and become frustrated when faced with lengthy documents''} find relief in AI’s ability to summarize these texts, as highlighted by P35. Furthermore, AI enhances study productivity by \textit{``creating practice tests from lecture slides, notes, and textbooks''}, which is particularly helpful for individuals with ADHD who struggle with organization while studying (P36).

\begin{table*}
    \resizebox{1.\linewidth}{!}{
        \renewcommand{\arraystretch}{1.45}
        \setlength{\tabcolsep}{4pt}
        \begin{tabular}{p{0.075\textwidth}p{0.2\textwidth}p{0.06\textwidth}p{0.82\textwidth}}
            \toprule
             &
             &
            \textbf{Poster ID} &
            \textbf{Quotes} \\
            \midrule

            \centering\multirow{9}{*}{\rotatebox{90}{\renewcommand{\arraystretch}{1}\begin{tabular}[c]{@{}c@{}}\Challenget{\textbf{Challenges}}\\\Challenget{\textbf{with LLM Use}}\\ \end{tabular}}} 
            
            & \multirow{1}[2]{0.2\textwidth}{\Challengecode{Prompting frustrations}} 
            & P40 & I find it challenging to write the right prompts to generate the responses I am looking for. (Autism) 
            \\
            \arrayrulecolor{buse_red!30!white}\cline{3-4}    
            & \multirow{1}[2]{0.2\textwidth}{\Challengecode{Neurotypical biases in LLM responses}} 
            & P41 & The effectiveness of AI is constrained by its training data. Most of the AIs trained on  internet, that is why AI' judgments are based on majority's perspective. Therefore, it cannot capture the thought process for people like me. This can be unhelpful for those with ADHD due to their divergent thinking since ADHD only affects a small part of the population. (ADHD) 
            \\
            \arrayrulecolor{buse_red!30!white}\cline{3-4}    
            & \multirow{1}[2]{0.2\textwidth}{\Challengecode{Lack of personal voice}} 
            & P42 & When I use ChatGPT for my professional writing, I need to adjust to maintain my authentic voice. (Autism)
            \\ 
            \arrayrulecolor{buse_red!30!white}\cline{3-4}    
            & \multirow{1}[2]{0.2\textwidth}{\Challengecode{Text- centric interactions}}  
            & P43 & We still need to read responses from ChatGPT, so I am not sure if it is helpful. My reading speed is already slow to average, and I often struggle to fully understand the content. (DYS) 
            \\
            \arrayrulecolor{black}\midrule

            \centering\multirow{9}{*}{\rotatebox{90}{\renewcommand{\arraystretch}{1}\begin{tabular}[c]{@{}c@{}}\Needst{\textbf{Needs and Wants}}\\ \end{tabular}}} 

            & \multirow{2}{0.2\textwidth}{\Needscode{Multimodal interactions}} 
            & P44 & I made this some time ago if you want to try it ahead of time. It adds the speaking/hearing functionality [to ChatGPT]. (DYS)   
            \\
            \arrayrulecolor{buse_seafoam!30!white}\cline{3-4}    
            & \multirow{2}{0.20\textwidth}{\Needscode{ND-friendly prompts}} 
            & P45 & How can I do this with ChatGPT? Because apparently, I have a cold face even when I am writing emails. Using ChatGPT to adjust my tone like this would be fabulous!  (Autism)
            \\
            \arrayrulecolor{buse_seafoam!30!white}\cline{3-4}    
            & \multirow{2}{0.2\textwidth}{\Needscode{LLM tools to support daily tasks}} 
            & P46 & An AI that functions as an executive coach would be very helpful. It can tell users about their daily goals and schedule tasks accordingly, reminding you to take breaks and to tell to switch between tasks. I find it nearly impossible to overcome inertia without external motivation. (ADHD)
            \\
            \arrayrulecolor{buse_seafoam!30!white}\cline{3-4}    
            & \multirow{2}{0.2\textwidth}{\Needscode{Greater acceptance of LLM use by ND users}} 
            & P47 & I believe students, especially those with dyslexia, should have access to AI tools. Avoidance is not a solution. These tools are not going anywhere and we should use them for our benefits rather than avoiding it. (DYS)
            \\
            \arrayrulecolor{black}\midrule
                        
            \centering\multirow{9}{*}{\rotatebox{90}{\renewcommand{\arraystretch}{1}\begin{tabular}[c]{@{}c@{}}\Sharet{\textbf{Sharing}}\\\Sharet{\textbf{Hacks \& Resources}}\\ \end{tabular}}} 

            & \multirow{2}{0.25\textwidth}{\Sharecode{Prompting hacks}} 
            & P48 & Check out this ChatGPT prompt for creating a personalized ADHD coach: 'My name is... You are my personal ADHD coach. Please ask me to list today’s tasks. After I share them, help me prioritize them based on time constraints and difficulty level, starting with the simplest tasks to build momentum for the more challenging ones. (ADHD)
            \\
            \arrayrulecolor{buse_mustard!30!white}\cline{3-4}   
            & \multirow{1}{0.20\textwidth}{\Sharecode{LLM applications for ND users}} 
            & P49 & Writing emails can be a nightmare due to all the unspoken neurotypical rules. Luckily, Grammarly now offers a ChatGPT-like AI that edits text to adjust different tones or goals like formal, casual, or assertive. It is like having a neurotypical friend review your seemingly harmless email to ensure it does not come across as unintentionally aggressive, without the discomfort of actually asking someone. (Autism)
            \\
            \arrayrulecolor{buse_mustard!30!white}\cline{3-4}   
            & \multirow{1}{0.20\textwidth}{\Sharecode{LLM applications built by ND users}} 
            & P50 & I was dealing with anxiety and have no one to talk to, so, I developed an AI chatbot that uses a method known as theory of mind to understand your emotions and needs, and respond to you appropriately. It adjusts to your reading style and preferences, serving as a personalized psychological partner. (SA)
            \\
            \arrayrulecolor{black}\midrule
            
            \centering\multirow{6}{*}{\rotatebox{90}{\renewcommand{\arraystretch}{1}\begin{tabular}[c]{@{}c@{}}{\textbf{\Concernt{Concerns}}}\\ \end{tabular}}} 
            
            & \multirow{1}{0.25\textwidth}{\Concerncode{False information}} 
            & P51 & Be cautious about learning about autism from AI without verifying the facts. AI can occasionally give incorrect info, and also its source of information is internet, which includes plenty of misinformation. (Autism) 
            \\
            \arrayrulecolor{buse_blue!30!white}\cline{3-4}    
            & \multirow{1}{0.20\textwidth}{\Concerncode{Replacing human connections}} 
            & P53 &  AI chatbots are not good for us. They mimic human interaction and might lead us to avoid real human interaction, resulting in increased isolation. (Autism)
            \\
            \arrayrulecolor{buse_blue!30!white}\cline{3-4}    
            & \multirow{1}{0.18\textwidth}{\Concerncode{Overreliance}} 
            & P54 & I worry that using ChatGPT might make me give up real friends, as it is much easier, but I plan to set limits for myself to prevent that from happening. (Autism) 
            \\
             
        \arrayrulecolor{black}\bottomrule
        \end{tabular}
    }
\caption{Challenges with LLM use, sharing needs and wants, sharing hacks and resources, and  concerns of using LLMs among neurodivergent individuals with corresponding paraphrased quotes from users across autism, social anxiety, ADHD, and dyslexia subreddit communities.}
\label{tab:rest_quotes}
\end{table*}

In addition, AI helps individuals with \PR{organizing train of thoughts}. P37, who has autism, appreciates ChatGPT for its capability in \textit{``structuring scattered ideas into a coherent format''} while P38, with dyslexia, similarly found value in using AI to think more deeply and straightforwardly about subjects. 

AI also serves as a \PR{career development resource}, especially during the job application processes. For individuals with ADHD, tailoring resumes and cover letters to fit specific job descriptions can be tedious and draining. P39 attests to the value of AI in this context, noting \textit{``AI has been so helpful in quickly adapting my resumes and cover letters to match job description''}. Identifying transferable skills is another area where AI can provide support, as P40 suggested: \textit{``ask[ing] for a list of transferable skills from past jobs to the position you are applying for''}.

\subsection{When interacting with LLMs, what challenges do neurodivergent individuals encounter?}
While LLM tools like ChatGPT provide various forms of assistance, neurodivergent users often face \Challenge{challenges} in using these tools effectively, as detailed in Table \ref{tab:rest_quotes}. Many users express \Challenge{prompting frustrations} as: \textit{``I find it challenging to write the right prompts to generate the responses [they are] looking for''} (P40). Moreover, a Redditor from the ADHD community noted how hard to achieve a desired response:  \textit{``I have been trying to do this same thing with ChatGPT but it gives me an unrelated response ... maybe my wording is wrong.''}

Besides prompting difficulties, neurodivergent users also face with \Challenge{neurotypical biases in LLM} \Challenge{responses}. Some, like a Redditor with ADHD, find that ``\textit{ChatGPT's outputs are very neurotypical, as expected, so they do not clearly capture my thought process, leading me through many rounds of prompting and refining responses.''} This issue arises because \textit{'most AIs are trained on the internet, which is why their judgments are based on the majority’s perspective'} (P41). As a result, ChatGPT may \textit{``not capture the thought process [of those with ADHD, for example]''} and is \textit{``less helpful [for them] due to their divergent thinking''} (P41).

Additionally, some users express frustration with ChatGPT's content moderation policies as another aspect of the biases, which can restrict discussions on certain topics like neurodiversity. These policies, while intended to protect users, can sometimes limit users from exploring and expressing their experiences with their conditions and hinder meaningful communication for those seeking support or understanding, as shared by a Redditor from the autism subreddit:

\begin{quote}
    \textit{ChatGPT has strict rules against discussing some topics like autism in negative terms. I often use it to search words, but occasionally it offers standard responses, stressing that it is inappropriate to discuss autism in certain ways and emphasizing the spectrum's variety in characteristics and behaviors. Rather than being helpful, it feels like these restrictions seem deliberately imposed to prevent specific conversations, which might be wise, yet I find it limiting and somewhat frustrating.}  (Autism)
\end{quote}

Neurodivergent users also express limitations of LLMs due to the \Challenge{lack of personal voice}. Although users acknowledge the utility of ChatGPT's responses, they often find these as lack of authenticity, as P42 points out, \textit{``I need to adjust [ChatGPT's response] to maintain my authentic voice''} indicating a need for personalization to maintain their own voice in the responses. This need for adaptation is further emphasized by another user who stated, \textit{``ChatGPT gave me something usable, but I had to modify the response.''} As a result, while the initial response provides a good starting point, users need to adjust the responses to ensure that the final message accurately represents their personal tone and style.

Furthermore, the \Challenge{text-centric interactions} pose a significant challenge for especially individuals with reading and writing difficulties, such as dyslexia. P43, for example, highlighted how the text-based nature of LLMs serves as a substantial barrier: \textit{``We still need to read responses from ChatGPT, so I am not sure if it is helpful''}, highlights the issue that, despite the potential advantages of using ChatGPT, relying on text as the primary mode of interaction does not align with their needs.

\subsection{When interacting with LLMs, what needs and preferences do neurodivergent individuals have?}
The challenges faced by neurodivergent users when using LLMs shape their \Needs{needs and wants}. The text-centric nature of interactions leads them to seek \Needs{multimodal interactions}. For instance, a dyslexic Redditor discusses their efforts to overcome this limitation by saying, ``\textit{I made this some time ago if you want to try it ahead of time. It adds the speaking/hearing functionality''} (P44). Another user from an autism community expressed excitement about the possibility of interacting AI with visuals: \textit{``I haven’t come across a visual prompt yet, but it would be fascinating to try one out''}.

The difficulty of prompting often leads neurodivergent users to search for \Needs{ND-friendly prompts} that better suit their cognitive differences. For instance, P45, Redditor from the autism community, wants to learn how to adjust the tone because people often told them \textit{``having a cold face when writing emails''}. Another user with ADHD seeks prompts to help with executive dysfunction: \textit{``I am curious if anyone has prompts for ChatGPT that can help with organization/ planning/ prioritizing tasks specifically?''}

However, writing effective prompts is only one technical aspect of the broader support sought by neurodivergent users. They often look for \Needs{LLM tools to support daily tasks}. For someone with ADHD, this might be an \textit{``AI that functions as an executive coach to tell your daily goals and schedule your tasks''} (46) or for individuals with autism, this could be an app that can help clarify vague sentences: \textit{``I’m curious if anyone knows of an app that can help translate what people mean when their sentences are unclear.''}

Beyond technical support, neurodivergent individuals seek \Needs{greater acceptance of LLM use} for them, particularly in bridging educational gaps. For example P47 with dyslexia emphasized the benefits of LLM use for dyslexic students: \textit{``I believe students, especially those with dyslexia, should have access to AI tools. ...These tools are not going anywhere and we should use them for our benefits rather than avoiding it.''} Another dyslexic user argues with people who are against using AI for neurodivergent students, stating a lack of knowledge about how people learn and process information differently, and highlights the usefulness of using AI-focused tools for people with dyslexia: 

\begin{quote}
     \textit{AI-driven tools help dyslexic people identify their errors and how to fix them if they use these tools correctly. Every time I see a counter-argument, it is made by people who are not aware of the fact that people learn and process differently. Certainly, having a voice in writing is nice. However, what is the point of having a voice if it is full of errors? Are these professors going to allow these errors for dyslexic students? (DYS)}
\end{quote}

\subsection{When interacting with LLMs, what hacks and resources do neurodivergent individuals share to navigate these challenges?} 

Neurodivergent communities on Reddit actively support each other by \Share{sharing hacks and resources}. These communities provide a space for users to exchange \Share{ prompting hacks} tailored to their unique needs. For example, P48, who has ADHD, described how to \textit{``create a personalized ADHD coach''} for listing and prioritizing daily tasks based on time and motivation, while another member from the social anxiety community shares strategies for how to use ChatGPT to start conversations by suggesting prompts like: \textit{“ChatGPT, help me create a creative conversation starter for [...]”} or \textit{“ChatGPT, I could use some help [...].”} In addition, members of these communities also demonstrate how to modify the existing prompts to be more neurodivergent-friendly, as shown by an autistic Redditor:

\begin{quote}
    \textit{There are many ChatGPT prompts online like this: }
    \begin{itemize}
        \item [-] \textit{``I want you to act as a life coach. You should ask me detailed questions about my life. Continue to ask questions until you have a full understanding of my personal and professional life... ''}
    \end{itemize}
    
    \textit{It sounds great, but one problem I have is that my brain does not like it when people ask me questions. It has nothing to do with questions feeling too invasive. My mind just goes super blank. I think it’s related to Asperger's. So I made one quick fix:}

    \begin{itemize}
    \item [-] [User shares an aspie-friendly version of the prompt]
    \end{itemize}
\end{quote}

These communities also post \Share{LLM applications for ND users}. P49, for instance, shared a ChatGPT-like AI tool to help with frustration of writing emails without being dependent on anyone by \textit{``adjust different tones, like formal, casual, confident, friendly, assertive''} for workplace communications. Other users from the autism community also shared AI tools for \textit{``having conversations''} or \textit{``checking your daily emotions''}.

Moreover, there are \Share{LLM applications built by ND users} that are shared in these subreddits. For example, P50, with social anxiety who shares their tool, stating, \textit{``I was dealing with anxiety and have no one to talk to, so I developed a chatbot that uses a method known as theory of mind to understand your emotions and needs, and respond to you appropriately''}. While they design these tools, they try to make them \textit{``personalized''} by \textit{``adjusting their reading style and preferences''} (P50), providing customized assistance for their unique experiences. Meanwhile, other users play ChatGPT's new ``GPT'' builder for provide assistance in various areas, such as clarifying written text to make them more direct and easier to understand: 

\begin{quote}
    \textit{I recently created a new GPT using ChatGPT's new 'GPT' builder and I get your feedback to see if it is useful. The concept is that you can input an instruction from work or a potential employer, and it will simplify the instruction, making it clearer and more straightforward by eliminating or emphasizing any underlying nuances or hidden meanings.} (Autism)
\end{quote}

\subsection{When interacting with LLMs, what concerns do neurodivergent individuals have?}
Neurodivergent users acknowledge the advantages of AI but also express several \Concern{concerns} about using LLMs. For instance, there are concerns about LLMs providing \Concern{false information}, particularly for their conditions. Users, such as P51, advise against relying solely on AI for learning about autism without verification, cautioning that \textit{``AI can occasionally give incorrect information...''}.

These individuals' concerns extend to potentially \Concern{replacing human connections} with AI. Some users worry that some neurodivergent individuals might prefer AI over human, since it tends to say exactly what they want to hear. This concern is highlighted by a Redditor from the autism subreddit: \textit{``AI should not replace the feeling of being seen by another human since it can say exactly the right psychological thing at the exact right time''}. This replacement could lead to \textit{``avoiding real human interaction, resulting in increased isolation''} (P53). 

In addition, some users have concerns about developing an \Concern{overreliance} on AI, particularly for human connection and social interaction. For example, P54 expresses concern that relying on AI for chatting might lead to abandoning real friendships because it eases social anxiety: \textit{``I worry that using ChatGPT might make me give up real friends, as it is much easier.''} There is also a worry that relying on AI for interaction could deteriorate further their social skills:

\begin{quote}
    \textit{I appreciate that people can use AI as a tool to overcome challenges, but I worry that relying on these tools could lead to further atrophy of already weak skills.} (Autism)
\end{quote}

Nevertheless, there is a determination to maintain control among users, with P54 noting, \textit{``I plan to set limits for myself to prevent that from happening.''}

\section{Discussion}

\subsection{Incorporating Neurodivergent Perspectives into LLM Development \& Design Implications}
Prior studies suggest that most LLMs do not integrate the perspectives of neurodivergent individuals \cite{killian2023knock}, as these powerful models are typically trained and built from vast volumes of online data predominantly authored by those who are neurotypical. NT-leaning biases in LLM responses, as noted by a user with ADHD \cite{oneill2023amplifying,bender2021on}, leading to prompting frustrations as a result of LLMs failing to understand or respond to the unique communication styles and preferences of neurodivergent users \cite{weidinger_2022,Zhao_2023}.

In response to these challenges, our findings show that neurodivergent users have expressed needs for more ND-friendly prompts that cater to their discursive preferences. Some share specific prompting strategies with their community while others even go further by modifying or building LLM-applications better suited for ND-specific needs. These hacks and exchanges of resources in relation to LLM interactions within neurodivergent communities are in line with prior research that shows how neurodivergent users often adapt ``Do-It-Yourself'' approaches \cite{hurst2011empowering,dix2007designing} in modifying existing technologies to overcome barriers. For instance, autistic players in Minecraft often modify their virtual environment in the game by creating ``sensory regulating spaces'' to accommodate sensory overloads \cite{ringland2016would}. Some argue that these adaptations reflect the potential additional burden on already marginalized communities \cite{edwards2023measuring}, stressing the importance of inclusive design in technologies. 

However, there are several potential challenges in making LLMs more inclusive for neurodivergent users. First, researchers have limited understanding of what constitutes neurotypical biases in LLM responses and what makes LLMs more ND-friendly. A key first step would be to learn from the communities for whom the LLMs are designed by engaging with neurodivergent users. Researchers in the past have suggested curating a diverse training data \cite{weidinger_2022} or offering multimodal LLM interactions \cite{zolyomi2023desinging,motti2019designing}. However, some of these solutions involve significant privacy risks and ethical concerns. For instance, fine-tuning LLMs using data from neurodivergent individuals raises concerns about privacy risks and potential misuse of sensitive or personal information \cite{brown_2022,sebastian2023privacy}, which are already major concerns for neurodivergent individuals \cite{valencia_2023}, who fear stigmatized views around their neurodivergent conditions, as well as outside perceptions of their dependence on LLMs, as shown in our findings.  

Building on these insights, our research offers several design implications to guide the development of more inclusive LLMs:

First, \textbf{incorporating different modalities} should be prioritized. Our findings reveal that neurodivergent individuals, particularly those with dyslexia, encounter significant challenges with text-centric design. Participants expressed frustration with the reliance on text, as one user noted, \textit{``We still need to read responses from ChatGPT, so I am not sure if it is helpful.''} Another user with dyslexia suggested integrating voice input and output features as a potential solution. These insights also align with previous research that identifies the difficulties of text-based interactions in LLMs \cite{botchu2024can} and advocates for the inclusion of other modalities, like speech recognition for improving the accessibility \cite{botha2024neurodiversity}. Therefore, we recommend that future designs incorporate different modalities to better accommodate users, especially with reading and writing difficulties.

Second, the need for \textbf{personalized interactions} emerged in our findings. Our findings showed that neurodivergent users often find LLM responses useful but lacking personal relevance which needs adjustments to reflect their communication style. For instance, users noted the necessity to modify ChatGPT's outputs to align with their \textit{``authentic voice''}. A similar emphasis on personalization is highlighted in prior research, particularly for users with diverse communication needs \cite{jang2024s}. To address this, we recommend that future LLM tools include more robust personalization features that can match better to their unique communication style.

Third is \textbf{incorporating neurodivergent perspective in model training} to reduce neurotypical bias LLMs responses. Current LLMs are predominantly trained on data that reflects neurotypical perspectives, often overlooking the communication styles of neurodivergent individuals \cite{bender2021on,lazaridou2020emergent}. Our findings also highlight this issue, as neurodivergent users noted that \textit{``ChatGPT’s outputs are very neurotypical,''}, resulting in responses that \textit{``do not clearly capture their thought process,''}, as one user from an ADHD subreddit pointed out. To address this, we suggest the inclusion of more diverse and representative training datasets that incorporate neurodivergent users' voices and experiences to reduce the biases in LLM responses.

Fourth, it is important to develop \textbf{ND-friendly prompts} that align with the diverse cognitive needs of neurodivergent users to improve their interaction with LLMs. Neurodivergent users often struggle to write prompts that generate useful responses from LLMs, as existing prompt structures are generally designed with neurotypical users in mind. For example, an ADHD user sought prompts that could assist with executive dysfunction, specifically asking for help with \textit{``organization, planning, and prioritizing tasks''}. Our findings revealed that, in response to these challenges, users often develop and share their own ND-friendly prompts within their communities. To address this need, we recommend that future LLMs include built-in ND-friendly prompt templates and guidance. These tools should be designed to accommodate the unique cognitive processing styles of neurodivergent users, making it easier for them to interact with LLMs and receive more relevant responses.

\subsection{Concerns of Overreliance and Stigma around LLM Use Among Neurodivergent Users}

Overreliance on LLMs is a significant concern shared by many LLM users beyond those who are neurodivergent \cite{kapania2024m}. Some argue that excessive reliance on persuasive, immediate responses from LLMs could lead to declining writing and communication skills \cite{FortuneChatGPT2023}. This holds especially true as users focus more on effective prompting rather than improving their own writing abilities. Some researchers argue that atrophy in such skills could be exacerbated by a decrease in social interactions, as people increasingly turn to LLMs rather than others to seek advice on  professional and personal matters \cite{Cook2023ChatGPT}. Our findings also reveal that overreliance and skill atrophy are major concerns among neurodivergent communities, especially for users from social anxiety and autism subreddits, with some raising specific concerns that  over-dependence on LLMs could potentially lead to losing real friends or decreasing their social skills.

Another concern for neurodivergent individuals is how others may perceive their use of LLMs. Previous research highlights that using assistive technologies has a risk of emphasizing the differences of neurodivergent individuals, thereby causing negative perceptions and hence disinclination to use these tools \cite{ijerph20105773,parette2004assistive}. Our findings align with this concern, as one Redditor from a social anxiety community was concerned about using AI for emotional and mental health support would make them appear \textit{``sad and pathetic''}. Similarly, another user from an autism community acknowledged, \textit{``This is embarrassing, but talking to ChatGPT would make me feel less lonely''}. Unsurprisingly, our analysis shows that many users across all four neurodivergent groups express a strong desire for greater public awareness and acceptance of their use of LLMs, especially in workplace and educational settings. 

Given the benefits of using LLMs as expressed by neurodivergent individuals, we as researchers are challenged to consider how we can reduce the stigma associated with the use of LLMs in professional, social, and personal contexts. We cannot change the negative perceptions overnight, but we can reframe our design goals as we develop LLM-powered tools to assist neurodivergent individuals. Most current approaches, as noted in previous research \cite{o2018risks,motti2019designing}, aim at changing the nature of neurodivergent individuals rather than adapting environments to better suit their needs. Instead, as suggested by O’Brolcháin (2018), we should focus on developing tools that serve as bridges between neurodivergent and neurotypical individuals \cite{o2018risks}, paralleling comments on Reddit that the role of LLMs needs to be seen as a ``mediator between neurodivergent and neurotypicals''. This might help clarify the role of these tools, similar to how other tools, like calculators, are everyday tools. As one dyslexic user described, LLMs are a ``writing version of a calculator,'' demonstrating just how effective the tool is at integration with the user's day-to-day activities and at assisting a variety of needs. We can make these tools less stigmatizing by adapting this perspective and encouraging their use to assist neurodivergent individuals.

\subsection{Balancing Interpersonal Communication Support and the Risk of Encouraging Masking in Neurodivergent Users of LLMs }

Bridging the gap between neurodivergent and neurotypical people requires effort from both sides \cite{Lowy2023Toward}, yet the burden often disproportionately falls on neurodivergent individuals \cite{milton2022double}. For instance, there are expectations that autistic people conform to neurotypical communication standards \cite{milton2022double}, exacerbating the double empathy problem \cite{milton2012ontological}. Similarly, many prior studies show that neurodivergent students are expected to adapt to traditional learning environments \cite{o2018risks}. These expectations can in turn shape how neurodivergent users use technologies, including LLMs.

Our findings show that neurodivergent individuals, especially those from the autism and social anxiety communities, frequently use LLMs for interpersonal communication support. Such users often use LLMs as a bridge in translating between neurodivergent and neurotypical perspectives (referred to as an ``NT-to-ND Translator'' by a user from autism communities) and to interpret social encounters with neurotypical peers. Some users from both communities use LLMs on a daily basis to convey appropriate tone and intention to others. Many rely on LLMs as conversational partners to rehearse discussions with neurotypical individuals that are anticipated to be difficult. Likewise, many ADHD users often use LLMs to help distill long conversations or to actively engage in extended conversations with their neurotypical peers by refining their replies to be focused and clear, avoiding looking like a fool in their conversations with neurotypical peers. 

As such, LLMs may provide immediate valuable interpersonal communication support, as expressed by the users themselves in our findings. However, such use cases also reflect a tendency among users to use LLMs to support their masking efforts. Masking refers to the process where autistic individuals consciously or unconsciously alter their behavior to conform to social norms or to hide characteristics of their autism from others \cite{pearson2021conceptual}. Research shows that frequent masking is associated with poor mental health outcomes, including depression, anxiety, and burnout among neurodivergent individuals \cite{Radulski2022Conceptualising}. While LLMs can be valuable tools in aiding interpersonal communication, there is an underlying risk that their use over time may inadvertently reinforce some users to adhere to neurotypical norms and expectations. As researchers, we must consider such implications of these technologies and explore ways to design technologies that avoid exacerbating the current burden of bridging efforts faced by neurodivergent individuals.

\section{Conclusion}
In this study, we analyzed posts from neurodivergent individuals' on Reddit to gain insight into how they engage with LLMs. We further examined the challenges they face, the workarounds they use to alleviate these difficulties, and their preferences and concerns around using LLMs. Our results revealed that neurodivergent individuals use LLMs across five main areas including Emotional Well-Being, Mental Health Support, Interpersonal Communication, Learning, and Professional Development and Productivity. We also identified key challenges, needs and wants, sharing hacks \& resources, and concerns. These findings indicate that neurodivergent and neurotypical users have different perspectives and expectations from LLM tools. This demand is further expressed as demanding more ND-friendly tools and interactions. Although neurodivergent users find existing tools useful for mitigating difficulties in their daily activities, some fear that overreliance on LLMs may further weaken their skills in areas where they struggle.

\textbf{Limitations and future work.} While our study provides insight into neurodivergent users' perceptions and engagement with LLMs, it has several limitations. First, our data collection was limited to Reddit, where members are mostly anonymous. As such, these findings may not accurately represent the views of a larger group of people. Second, we rely on users' self-reported statements to be truthful, which might not always be accurate. Third, there is only one subreddit dedicated to dyslexia and social anxiety. This prevents us from investigating how these groups interact on various subjects. Further research is needed to explore these results by reaching out to neurodivergent individuals to gain insights directly from them. Exploring how neurodivergent individuals perceive and use LLM tools can yield significant insights that will help in the development of more inclusive and accessible LLM tools.

\begin{acks}

\end{acks}

\bibliographystyle{ACM-Reference-Format}
\bibliography{all, Neurodiversity}

\appendix
\clearpage

\section{Neurodivergent Conditions}

\begin{table}[!hb]
    \centering
    \begin{tabular}{ll}
    \toprule
        Asperger syndrome & Attention Deficit Hyperactivity Disorder \\ 
        Auditory Processing Disorder & Autism spectrum disorder \\ 
        Bipolar disorder & Central Auditory Processing Disorder \\ 
        Developmental Language Disorder & Down’s Syndrome \\ 
        Dyscalculia & Dysgraphia \\ 
        Dyslexia & Dyspraxia \\ 
        Hyperlexia & Hypersensitivity \\ 
        Hyposensitivity & Intellectual disabilities \\ 
        Irlen Syndrome & Misophonia \\ 
        Non-Verbal Learning Disorder & Obsessive-compulsive disorder \\ 
        Prader-Willi syndrome & Schizophrenia \\ 
        Sensory processing disorders & Social anxiety \\ 
        Stammering & Tourette’s syndrome \\ 
        Williams syndrome & \\
        \bottomrule
    \end{tabular}
    \caption{Comprehensive list of neurodivergent conditions.} 
    \label{conditions}
\end{table}

\section{Reddit Communities}
\begin{table}[ht]
    \centering
    \begin{tabular}{lllll}
    \toprule
    /r/ADHDers & /r/ADHDmeme & /r/ADHDmemes & /r/ADHDmeme \\
    /r/adhd\_anxiety & /r/adhdwomen & /r/adultautism & /r/aretheNTsokay \\
    /r/Aspie & /r/asperger & /r/aspergers & /r/aspergirls \\
    /r/aspergirls & /r/aspiepositivity & /r/AskAutism & /r/AskAutism \\
    /r/AuDHD & /r/AuDHD & /r/AuDHDWomen & /r/auDHD \\
    /r/AutisticAdults & /r/AutisticCreatives & /r/AutisticPOC & /r/AutisticPride \\
    /r/AutisticWithADHD & /r/Autism & /r/autism & /r/Autism\_Parenting \\
    /r/AutismInWomen & /r/AutismTranslated & /r/autismafterdark & /r/autisticpeeps \\
    /r/autism & /r/Aspergers & /r/AutismTranslated & /r/BisexualswithADHD \\
    /r/bipolar & /r/Dyslexia & /r/dyscalculia & /r/dysgraphia \\
    /r/dyspraxia & /r/hyperlexia & /r/hypersensitivity & /r/irlensyndrome \\
    /r/LearningDisabilities & /r/LGBTaspies & /r/misophonia & /r/neurodiversity \\
    /r/neurolationships & /r/OCPD & /r/schizophrenia & /r/socialanxiety \\
    /r/SPD & /r/spicyautism & /r/Tourettes & /r/TwoXADHD \\
    /r/TwoXADHD & /r/williamssyndrome & & \\
    \bottomrule
    \end{tabular}
    \caption{List of subreddits related to neurodivergent conditions.} 
    \label{communities}  
\end{table}

\end{document}